\documentstyle[sprocl,psfig,epsf,amssymb]{article}

\def\Journal#1#2#3#4{{#1} {\bf #2}, #3 (#4)}

\def\NPB{{\em Nucl. Phys.} B}
\def\NPA{{\em Nucl. Phys.} A}
\def\PLB{{\em Phys. Lett.}  B}
\def\PRL{\em Phys. Rev. Lett.}
\def\PRD{{\em Phys. Rev.} D}
\def\PRC{{\em Phys. Rev.} C}
\def\PRB{{\em Phys. Rev.} B}

\def\PREP{\em Phys. Rept.}
\def\IJMPA{{\em Int. J. Mod. Phys.} A}

\def\be{\begin{equation}}
\def\ee{\end{equation}}
\def\bea{\begin{eqnarray}}
\def\eea{\end{eqnarray}}

\begin{document}

\title{THE QCD TRICRITICAL POINT: BEYOND MONOTONY 
IN HEAVY ION PHYSICS\footnote{Much of this work 
was done in 
collaboration with M. Alford
and F. Wilczek,\cite{arw1,3flavor} with J. Berges \cite{bergesraj}
and with M. Stephanov and E. Shuryak.\cite{signatures} \hfill Preprint
MIT-CTP-2774.}}

\author{KRISHNA RAJAGOPAL}

\address{Center for Theoretical Physics, MIT, Cambridge, MA 02139\\
E-mail: krishna@ctp.mit.edu}

\maketitle

\abstracts{I first sketch recent developments concerning the phase
diagram of strongly interacting matter as a function of temperature  
and baryon density, obtained using a model for two-flavor QCD in 
which the interaction between quarks is modelled on that induced by 
instantons.  The phase diagram has a chiral symmetry breaking vacuum, a color 
superconductor phase at high density, and a mixed phase in between which can 
describe nuclear matter if the high density regions in the mixed phase form 
droplets. This is amusingly similar to the phase diagram of the copper oxide 
superconductors, although their critical temperatures are smaller by a factor 
of about 10$^{-9}$ and their mixed phase seems to form stripes, rather than 
droplets.  This and other approaches to QCD with two {\em massless} quarks 
suggests the existence of a tricritical point on the boundary of the phase 
with spontaneously broken chiral symmetry.  In QCD with {\em massive} quarks 
there is then a critical point at the end of a first order transition line.  
We discuss possible experimental signatures of this point, which provide 
information about its location and properties.  We propose a combination of 
event-by-event observables, including suppressed fluctuations in $T$ and $\mu$
and, simultaneously, enhanced fluctuations in the multiplicity of soft pions.
As a control parameter (like the collision energy) is varied, the fluctuations
of appropriate event-by-event observables should therefore have minima or 
maxima.  Experimental detection of these non-monotonic signatures would 
confirm that the matter was initially above the chiral phase transition, 
increasing our confidence in the interpretation of monotonic changes in other 
observables, and would teach us much about the QCD phase diagram.}

\section{Introduction}

In QCD with two massless
quarks, a spontaneously broken chiral symmetry is restored at 
finite temperature. It can be argued \cite{piswil,rajwil,rajreview}
that this phase transition is likely second order 
and belongs to the universality class of $O(4)$ spin models in 3 dimensions.
If this transition is indeed
second order, QCD with two quarks of nonzero mass has
only a smooth crossover as a function of $T$. 
Although not yet firmly established, this picture is consistent
with present lattice simulations  and many models. 

At zero $T$ several models 
suggest \cite{njl,barducci,steph,arw1,rssv,bergesraj,stephetal} that
the chiral symmetry restoration transition at finite $\mu$ is {\it first}
order.  Assuming that this is the case in QCD, one can easily argue
that there is a tricritical point in 
the $T\mu$ phase diagram.\cite{barducci,bergesraj,stephetal}
In QCD with massless quarks,
there is a sharp boundary in the $T\mu$ plane separating
regions with chiral symmetry manifest from those in which
it is broken.   The point on this boundary at which the transition
changes from first to second order is by
definition tricritical.
The nature of this point can be understood
by considering the Landau-Ginzburg effective potential for the
order parameter of chiral symmetry breaking, 
$\phi=(\sigma,{\pi})\sim\langle\bar\psi\psi\rangle$:
\begin{equation}
\Omega(\phi) = a\phi^2 + b(\phi^2)^2 + c(\phi^2)^3.
\label{phi6}
\end{equation}
The coefficients $a$, $b$, and $c>0$ are functions of $\mu$ and $T$.
The second order phase transition line described by $a=0$ at $b>0$
becomes first order when $b$ changes sign. The critical properties
of this point can be inferred from universality,\cite{bergesraj,stephetal}
and are characterized 
by the exponents of the mean field theory
(\ref{phi6}), and
calculable logarithmic corrections to 
scaling,\cite{lawrie} because the critical dimension
above which $\phi^6$ theory has mean field exponents is $d=3$.
If two-flavor QCD has a second order transition at
high temperatures and a first order transition at high
densities, then it will have a tricritical point in
this universality class.

Away from the chiral limit, the second order chiral transition
turns into a smooth crossover, while the first order
transition remains first order. Of particular
interest is the fact that the tricritical point
becomes an Ising second order transition.\cite{bergesraj,stephetal}
This raises the possibility that even though the pion
is massive in nature, long correlation lengths 
and divergent susceptibilities may arise
in heavy ion collisions which traverse the chiral 
transition with a suitable chemical potential. 
We describe the resulting non-monotonic signatures \cite{signatures}
in Section 4. In Section 2, we first sketch one of the 
models \cite{bergesraj} within which the tricritical point
arises explicitly.  The color superconducting phase of 
QCD can also be studied within this model, and we note
several analogies between
the phase diagram which emerges and that
of the cuprate superconductors. In Section 3 we discuss the effect
of the strange quark on the position of the tricritical point.

\section{A Model and an Amusing Analogy}

\begin{figure}[thb]
\centerline{
\epsfysize=10cm
\epsfbox[81 202 487 717]{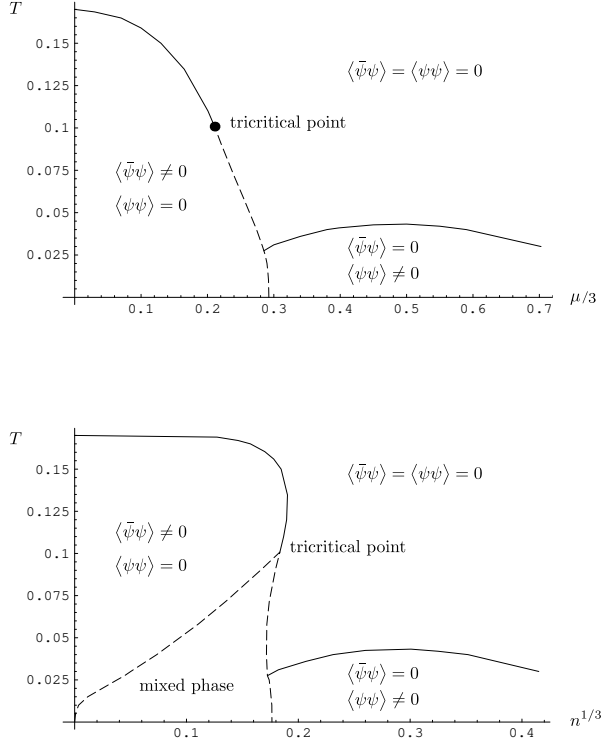}
}
\vspace{-0.1in}
\caption{Phase diagram as a function of $T$ and $\mu/3$,
and as a function of $T$ and $n^{1/3}$ (all in GeV) 
for the 
model for QCD with two massless quarks
described in Ref.$^1$.  This section
can be viewed as one long caption for these figures. The
solid curves are second order phase transitions; the dashed curves
describe the first order phase transition.}
\end{figure}
Berges and I have done a qualitative exploration of the phase
diagram of two-flavor QCD as a function of temperature and 
chemical potential
in a context which allows us to describe
likely patterns of symmetry breaking and to make rough
qualitative estimates.\cite{bergesraj} As a tractable model, we consider 
a class of fermionic models for QCD where the quarks
interact via a four-fermion interaction modelled on 
that induced by instantons which respects all the symmetries
of QCD.\cite{arw1}
There are two
competing ordering possibilities in 
QCD at nonzero density, and both
can be studied within this model.  At low temperatures
and chemical potentials, we expect chiral symmetry
breaking via a nonzero chiral 
condensate $\phi\sim\langle\bar\psi\psi\rangle$.
At large chemical potentials and low temperatures,
we expect the quark Fermi surfaces to be unstable
to the formation of a condensate of diquark Cooper pairs, leading
to a superconductor in which gauge symmetries are broken
and there is a gap $\Delta\sim\langle \psi \psi \rangle$ 
in the spectrum of fermionic
excitations.\cite{barrois,bailin,arw1,rssv,3flavor,bielefeld}  
In Ref. \cite{bergesraj}, the 
model is presented in detail and choices of parameters
are discussed.  We evaluate the thermodynamic potential
as a function of $\phi$ and $\Delta$ for different $T$ and
$\mu$, and obtain the phase diagram shown in Fig. 1.

At zero temperature, we find a first order phase transition
between a phase with density zero in which chiral symmetry
is broken, namely the vacuum, and a phase with density 
$n_0$ in which chiral symmetry is restored.  
This has an interesting interpretation.\cite{buballa,arw1,bergesraj}
For reasonable choices of parameters, $n_0$ is greater than
nuclear matter density and comparable to the baryon density
in a {\it nucleon}.  Ordinary nuclear matter
is then in the mixed phase of this transition and consists
of nucleon-sized droplets within which $n=n_0$ and $\phi=0$,
surrounded by regions of vacuum within which $n=0$ and
$\phi\neq 0$.   The model as presently
analyzed does not give a complete description of nuclear 
matter.  First, color must be gauged if the model is to
yield droplets which are color singlets.  Second, a short
range repulsion and long range attraction between droplets must be included 
so that small droplets are favored over bigger ones,
and so that the droplets attract each other to form nuclei.
Were one to add these interactions, one could describe
the liquid-gas transition (beween the nuclear matter liquid
and a low density gas of nucleons) observed in low
energy nuclear collisions,\cite{csernai}
as discussed recently.\cite{stephetal}

Once nuclear matter is squeezed enough that
the transition to the phase with density $n_0$ 
and above is completed, one obtains
a spatially homogeneous color superconductor.  
The Cooper pairs which condense are color $\bar 3$, flavor
singlet, Lorentz scalars.  
We find critical temperatures
of order 25-40 MeV for the parameters in Fig. 1, 
and up to $\sim 100$ MeV for other choices of parameters.\cite{bergesraj}
The gap $\Delta$ is about 1.7 times
$T_c$.
Only two of the three colors participate in the condensate, meaning
that only 2/3 of the quark excitations at the fermi surface
have a gap. The phase transition at $T_c$ at which superconductivity
is lost is second order in our mean field analysis, but may become
either first order or a crossover when fluctuations are taken
into account.

In QCD with three massless
quarks, a phase diagram similar to that in Fig. 1 is plausible,
with two qualitative changes:  all the transition lines are
likely first order as we discuss below, and the symmetry properties of the
superconducting phase are strikingly different than in
the two flavor theory.\cite{3flavor}
The quark pair condensate which forms
at densities high enough that chiral symmetry is
restored must break flavor symmetries, since two quarks
cannot form a flavor singlet.  In fact, the condensate
which seems to be favored \cite{3flavor} ``locks'' color and
flavor rotations,
breaking the color
and flavor SU(3)$_{\rm color} \times$SU(3)$_L \times$SU(3)$_R$
symmetries down to the diagonal subgroup SU(3)$_{{\rm color}+L+R}$.
All quarks have a gap, and
therefore if this condensate occurs within neutron
stars, direct neutrino emission is blocked.
All gluons get a mass. There is an octet of pseudoscalar
approximate Nambu-Goldstone boson excitations of
the diquark condensate  reflecting the fact
that chiral symmetry is broken by color-flavor locking,
even in the absence of a $\langle \bar \psi \psi \rangle$ 
condensate. There is a modified but still massless photon;
with respect to this photon, one finds only integrally
charged excitations.  The thermal properties are dominated
by a neutral superfluid and by the Nambu-Goldstone excitations.
The gaps characterizing this phase are of order 10-100 MeV.\cite{3flavor}
Including a nonzero strange quark mass $m_s$ results in a mismatch
between the Fermi momenta for the light and strange quarks
given by $\mu - \sqrt{\mu^2-m_s^2}$, of order 10 MeV for $\mu\sim 500$ MeV.
As long as the gap is larger than this mismatch,
the strange quark mass should have only a quantitative effect.
Were the gap smaller, one would return to the two-flavor 
condensate described above, with the light and strange quarks
behaving independently.\cite{bailin}

The lower panel of Fig. 1 looks similar to the phase diagram
which describes the cuprate superconductors.\cite{birgeneau}
There, one can vary both the temperature and, by doping, the number density 
of holes in the copper oxide planes.  
At low densities, and at temperatures 
up to many hundred Kelvin, these materials are antiferromagnets.
Antiferromagnetic order and chiral symmetry breaking are similar.
Indeed,
the development of antiferromagnetic order can be mapped onto chiral
symmetry breaking in a 2+1 dimensional theory of Dirac fermions
and gauge fields.\cite{patrick}
Like QCD, and unlike a ferromagnet, an antiferromagnet has
Goldstone bosons with
a dispersion relation 
$\omega \sim k$.  
At higher densities and lower temperatures,
these materials are superconductors.\footnote{$T_c$ in these
materials can be over 
100 K; hence they are called 
high temperature superconductors.  However, as in 
Fig. 1, $T_c$ is 
still significantly lower than the temperature at which the antiferromagnetic
order is lost at zero density. 
$T_c$
for the QCD superconductor is many orders of magnitude higher
than that for any cuprate superconductor, 
but this is an unfair comparison! It is more appropriate to note
that in QCD we find $T_c$'s of order 10\% of the Fermi energy,
reasonable for a BCS superconductor at reasonably strong coupling.}  
In between the low density antiferromagnetic and 
high density superconducting phases,
there is a range of densities in which one finds a spin
glass, which has no obvious analogue in the QCD phase diagram.
However, if these materials were fully 3-dimensional, rather than
behaving 2-dimensionally in many ways,
the antiferromagnetic phase would likely extend
to higher densities.  

There may be a further analogy between QCD and the cuprate 
superconductors.
There
are indications that on the low density side of the
superconducting region in the cuprate phase
diagram, there may be a phase characterized by alternating antiferromagnetic
and superconducting stripes.\cite{stripes}
In the mixed phase in QCD, the
high density regions organize themselves into small nucleon-sized
droplets, while
in the copper oxide planes, it seems that the hole-rich
regions may  organize 
themselves into long stripes a single lattice spacing 
wide. As in QCD, this can be seen as a competition
between phase separation which reduces the disruption of the
magnetic (chiral) order by segregating the fermions, 
and repulsion between
the fermions which tries to scatter them. 
For whatever reason, this competition
results in narrow stripes in one case and small droplets in the
other.
Whereas the droplets in the QCD mixed phase are
so small (quark number 3) that they cannot be 
superconductors in any meaningful sense, the stripes
in the cuprates can be superconducting.

QCD and the cuprate
superconductors are both strongly interacting systems with
chiral symmetry breaking/antiferromagnetism
at low densities  and
superconductivity at higher densities and relatively low temperatures.
Furthermore, it may be that in both systems these phases 
are separated by complicated regions
of the phase diagram
describable as mixed phases: nuclear matter in one case, stripes
in the other.  The different microscopic Hamiltonians and, likely
most important, the different dimensionalities make it unlikely
that these analogies can be made more than amusing.  Perhaps, though,
qualitative ideas from one system could
prove useful in the study of the other.

\section{The (Tri)critical Point....}

\begin{figure}[htb]
\vspace{-0.1in}
\centerline{\psfig{file=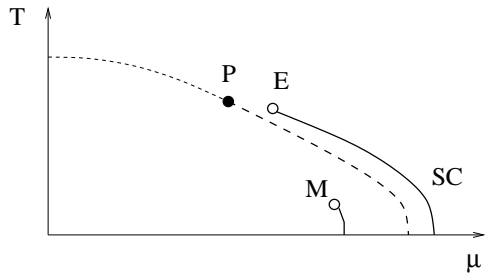,width=2.5in}}
\vspace{-0.1in}
\caption[]{The schematic phase diagram of QCD. The dashed lines
represent the boundary of the phase with spontaneously broken
chiral symmetry in QCD with 2 massless quarks. The point P
is tricritical. The solid line with critical end-point E
is the line of first order transitions in QCD with 2 quarks
of small mass. The point M is the end-point of the
nuclear liquid-gas transition probed in multifragmentation
experiments. The superconducting phase of QCD, 
marked SC, is not relevant to the rest of our discussion.} 
\label{fig:pd}
\end{figure}
We now set aside models and analogies, and return to our general
discussion of the tricritical point in the QCD phase diagram.
This section and the next describe work done with
M. Stephanov and E. Shuryak.\cite{signatures}
In real QCD with nonzero quark masses, the second
order phase transition becomes a
smooth crossover and the tricritical point becomes
a critical (second order) end-point of a 
first order phase transition line.
Universality arguments \cite{lawrie,stephetal}
also predict that the end-point E in QCD with small quark masses
is shifted with respect to the tricritical point P towards larger $\mu$ as 
shown in Fig. \ref{fig:pd}.
It can also be argued \cite{bergesraj,stephetal}
that the point E is in the universality class of the 
Ising model in 3 dimensions, 
because the $\sigma$ is the
only field which becomes massless at this point. (The pions
remain massive because of the explicit chiral symmetry breaking
by quark masses.) 
In this paper we discuss experimental
signatures of this critical end-point, which is in the
same universality class as that in the standard liquid-gas 
phase diagram.

The position of the points P and E in two-flavor QCD was estimated 
using two different models (that of Ref. \cite{bergesraj}
and a random matrix model \cite{stephetal})
as $T_P\sim 100$ MeV
and $\mu_P\sim 600-700$ MeV.
These are only crude estimates, since they are based
on modeling the dynamics of chiral symmetry breaking only.

The third (strange) quark has an important effect on the position
of the point P and, therefore, of the point E. At
$\mu=0$, if the strange quark mass $m_s$ is
less than some critical value $m_{s3}$, 
the second order finite $T$ transition becomes first
order.  This leads to a 
tricritical point in the $Tm_s$ plane.\cite{rajwil,ggp,rajreview}  
Theoretically, the origin of this
point is similar to the one we are discussing. In terms
of eq. (\ref{phi6}) the effect of decreasing $m_s$
is similar to the effect of increasing $\mu$: the coefficient $b$
becomes negative.  
What is important is that,
unlike $m_s$,
$\mu$ is a parameter which can be {\em experimentally} varied.

Clearly, the physics of the $T\mu$ plane is as in Fig. \ref{fig:pd} only
for $m_s>m_{s3}$. For $m_s<m_{s3}$, the transition
is first order already at $\mu=0$, and, presumably, remains first order
at all nonzero $\mu$.\cite{hsu} 
As $m_s$ is reduced from infinity, the tricritical point P of Fig. \ref{fig:pd}
moves to lower $\mu$ until, at $m_s =m_{s3}$, it reaches
the $T$-axis and can be identified with the tricritical
point in the $Tm_s$ plane.   The two tricritical points are 
continuously connected.
We assume that $m_s>m_{s3}$ which is
consistent with the lattice studies of Ref. \cite{columbia}. What is important
for us is that the qualitative effect of the strange
quark is to reduce the value of $\mu_P$, and thus of $\mu_E$,
compared to that in two-flavor QCD, since $\mu_P =0 $ at
$m_s=m_{s3}$.  This
shift may be significant, since lattice studies show that
the physical value of $m_s$ is of the order of $m_{s3}$.

Analysis of particle abundance ratios observed in central heavy ion collisions
\cite{stachel} indicates that chemical freeze-out happens near the
phase boundary, at a chemical potential $\mu\sim 500-600$ MeV at
the AGS (11 GeV$\cdot A$), while at the SPS ($160-200$ GeV$\cdot A$) it
occurs at a significantly lower $\mu \lesssim 200$ MeV.  In view of the effect
of the strange quark just discussed, the estimated position of P and E
\cite{bergesraj,stephetal} should be shifted from
$\mu_E\sim 600-700$ MeV to lower $\mu$. Thus, it may well be
between the SPS and the AGS values of $\mu$, 
and therefore the point E may be accessible at lower energy or
non-central collisions at the SPS.

\section{.... and its Signatures}

The strategy for finding the point E 
which we propose is based on the fact that this point
is a genuine critical point. 
Such a point is characterized
by enhanced long wavelength fluctuations of the order parameter which lead to
singularities in all thermodynamic observables.
In the liquid-gas phase transition in water, 
the observation of critical opalescence is 
perhaps the easiest way to detect these
unique properties characterizing physics near the critical point.
The signatures we propose can
play an analogous role in QCD.

It is important to have control parameters which can be adjusted
so that the system explores different points on the phase diagram.
For example, by increasing
the energy of the collision  one increases the initial
$T$ and decreases the initial $\mu$.  
A qualitatively similar
effect may be achieved by increasing centrality.
We will generically call the
control parameter which is varied ``$x$'', and take increasing
$x$ to mean increasing collision energy or 
increasing centrality. 
Scanning in centrality will 
almost certainly be the easiest, since in any given run
events with all impact parameters are present. 
However, comparison of AGS and SPS results demonstrates that
scanning in energy will yield a large variation in the $\mu$
at which the transition is crossed, whereas this
has not been demonstrated for scanning in centrality, which
may only provide fine tuning.
\begin{figure}[htb]
\centerline{\psfig{file=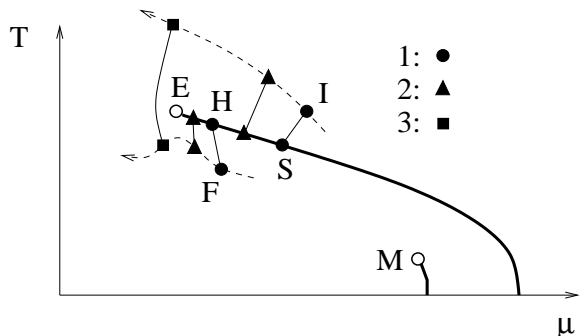,width=3in}}
\vspace{-0.1in}
\caption[]{Schematic examples of three possible trajectories for
three values of $x$ on the phase diagram of QCD
(see. Fig. \ref{fig:pd}). 
The points I, S, H and F on different trajectories
are marked with different symbols. The dashed lines show
the locations of the initial, I, and final, F, points
as $x$ is increased in the direction shown by the arrows.}
\label{fig:traj}
\end{figure}

In this work we do not discuss initial equilibration and we choose to
define the initial point, I$(x)$, as the point at which compression
has ended, most of the entropy is already produced, and approximately
adiabatic expansion begins.  The system will then 
follow some trajectory in the $T\mu$ plane
characterized by the ratio of the baryon charge density to the entropy 
density, $n/s$, which is (approximately) conserved.  Three trajectories
are shown schematically in Fig. \ref{fig:traj}. (For realistic 
hydrodynamical calculations
and discussion see, e.g., 
Refs. \cite{Frankfurt_hydro,Hung_Shuryak}).

Recall that the first order line in the $T \mu$ plane is actually a whole
region of mixed phase, with the additional hidden parameter being the
volume fraction of the two coexisting phases.
The  zig-zag shape occurs
because, as the trajectories enter the mixed phase region,
$n/s$ remains constant during the adiabatic expansion, 
latent heat is released and $T$ {\em increases}.
In
Fig. \ref{fig:traj}, we use the following notation: S$(x)$ 
for the ``softest'' point, H$(x)$ for
the ``hottest'' point and  F$(x)$ for the final thermal  
freeze-out after which no scattering occurs. 
(Note that at small values of $x$, 
at which the transition is first order, the trajectories
are likely to begin within the mixed phase region.
The special case when I$(x)$ coincides with S$(x)$ leads to
a local maximum
of the QGP lifetime \cite{HS_prl,Rischke}.)
Increasing $x$ will yield trajectories shifted to the left in 
Fig. \ref{fig:traj},
traversing the transition region at lower $\mu$ and
higher $T$.

The existence of the end-point singularity, E, leads to the 
phenomenon which we refer to as the ``focusing'' of 
trajectories towards E.   
The initial point I$(x)$ and the beginning of the zig-zag S$(x)$
depend on the control parameter $x$ more strongly 
than the zig-zag end-point H$(x)$.
The reason for this is that the point H$(x)$ is always closer to E than
S$(x)$ (see Fig. \ref{fig:traj}).  
%
%
This focusing effect implies
that exploring physics in the vicinity of the end-point singularity
may not require a fine-tuned $x$.
This situation resembles that in low energy nuclear collisions,
in which the first order liquid-gas phase transition also has a critical
end-point at a temperature
of order 10 MeV \cite{csernai} (point
M on Figs. \ref{fig:pd} and \ref{fig:traj}).   
In such experiments,  one varies control 
parameters
to maximize the probability of multi-fragmentation.
It was noticed long ago \cite{csernai} how
surprisingly easy it is to hit the critical region.
We believe that part of the reason is a focusing phenomenon 
analogous to the one we are describing.

Another aspect of the ``focusing'' 
arises via the divergence of susceptibilities, such as the
specific heat capacity $c_V=T\partial s/\partial T$, at the endpoint
E. As a result, the trajectories which pass near the critical point will
linger there longer.  
This makes it likely
that final freeze-out occurs at a temperature quite close to $T_E$,
rather than below it. So, while scanning in some control parameter $x$
and measuring the positions of the points F$(x)$, we may expect to find a
bump in the vicinity of the point E. (See the lower dashed curve on
Fig. \ref{fig:traj}.) 
At this point it is instructive to consider the dependence on another
control parameter, the atomic weight $A$ of the colliding nuclei.
If $A$ were infinite, the point F would be close to 
$T_F=0,\mu_F=m_N$. Thus, for $A$ large enough, the dotted curve 
in Fig. \ref{fig:traj} moves down and the bump 
and indeed all the signatures described below fade away. 
Experimentally, the $A$ dependence of the point $F$ has been 
established recently by the analysis of flow,\cite{Hung_Shuryak}
Coulomb effects\cite{Heiselberg} and pion interferometry.\cite{Heinz}
For example, in central S+S collisions at SPS $T_F\approx 140-150$
MeV, while for Pb+Pb it is only $T_F\approx 120$MeV.

We shall now discuss the signatures which
directly reflect thermodynamic properties of the system
near its critical point.
With the advent of wide-solid-angle detectors like NA49 at CERN,
it is now possible  to make {\em event-by-event}
measurements of observables which are proxies for the
freeze-out $T$ and $\mu$.\cite{NA49} 
We argue that the event-by-event fluctuations in 
both quantities should be anomalously small for 
values of $x$ such that the system freezes out near the critical
point.  As has been suggested recently,\cite{stodolsky,shuryak}
event-by-event fluctuations of $T$ 
can be related by basic thermodynamics to
the  heat capacity at freeze-out
\begin{equation}
\frac{(\Delta T)^2}{T^2} = {1\over C_V} \ .
\label{fluctuations}
\end{equation}
The quantity $C_V$ is extensive, so $\Delta T \sim 1/\sqrt{N}$ as expected,
where $N$ is the number of particles in the system.
If the specific heat $c_V$ diverges,  the coefficient
of $1/\sqrt{N}$ vanishes and fluctuations of $T$ are suppressed.
For freezeout in the crossover region, or in the hadronic
phase just below the first order transition, $c_V$ is 
finite. (If freeze-out occurs from the mixed phase,
some linear combination of the two susceptibilities is relevant.)
As the freeze-out point approaches the critical point
from either the left or the right,
$c_V$ diverges and $\Delta T \sqrt{N}$ is minimized.
Other susceptibilities, in particular, 
$-\partial^2 \Omega/\partial \mu^2$, are also divergent.
This implies that fluctuations of $\mu$ are also suppressed at the
critical point. Experimentally, $\Delta T$
can be found via event-by-event analysis of 
$p_T$ spectra.\cite{stodolsky,shuryak}
Fluctuations in
$\mu$ correspond to event-by-event fluctuations in the 
baryon-number-to-pion ratio.
The event-by-event fluctuations in experimental observables will receive
contributions in addition to the thermodynamic ones
we describe.  For example, the fluctuations in the
slope of the $p_T$ spectrum will receive a contribution
from $\Delta T$ and a (likely small \cite{shuryak}) contribution
from fluctuations in the flow velocity.
We therefore expect that as the collision
energy is increased so that the freeze-out point 
moves from right to left past the critical point, 
we will find minima (but not zeroes) in the widths
of the distributions of event-by-event
observables which are well-correlated
with $T$ and $\mu$.

Using universality, we can predict the exponents for the divergent
susceptibilities at the point E. Very naively, one might think that
the exponent describing the divergence of $C_V$ is $\alpha$, which is
small for the 3-dimensional Ising model universality class:
$\alpha\approx 0.12$.  In fact the exponent for $C_V$ is significantly
larger. This and the exponent for the $\mu$-susceptibility are
determined by finding two directions, temperature-like and
magnetic-field-like, in the $T\mu$ plane near point E,
following the standard procedure for mapping a liquid-gas
transition onto the Ising model.\cite{tsypin} 
The two linear combinations of $T-T_E$ and $\mu-\mu_E$
corresponding to these directions should then be identified (in the
sense of the universality) with the temperature, or $t=T-T_c$, and the
magnetic field, $h$, in the Ising model. 
The $t$-like direction should be tangential to
the first-order line at the point E.  Then $C_V$ and $-\partial^2
\Omega/\partial \mu^2$ are different linear combinations of the
$t$-like and $h$-like susceptibilities. In both linear combinations,
the divergence of the $h$-like susceptibility will dominate because
$\gamma\approx1.2 \gg \alpha\approx0.12$.  The exponent for the
divergence of the $h$-like susceptibility as a function of the
distance, $\ell$, from the point E will depend on the direction along
which one approaches this point. For almost all directions it will be
given by $\gamma/\beta\delta\approx0.8$ (except for exactly the $t$-like
direction, where it is $\gamma$).  As a result, for points on the
$T\mu$ plane along a generic line through E one finds
\begin{equation}
(\Delta T)^2 \sim (\Delta \mu)^2 \sim \ell\,^{0.8} 
\end{equation}
sufficiently close to E.
Therefore, the fluctuations of $T$ and $\mu$ are considerably 
suppressed when the freeze-out occurs near the 
critical point.

We turn now to direct signatures of the long-wavelength
fluctuations of the massless $\sigma$ field.
For the choices of
control parameters $x$ such that freeze-out occurs at (or near)
the point E, the $\sigma$-meson is the most numerous species
at freeze-out,
because it is (nearly) massless and so 
the equilibrium occupation number of the long-wavelength modes
($T/\omega$) is large.  Because the pions are massive at the critical
point E, the $\sigma$'s cannot immediately decay into $\pi\pi$.
Instead, they persist as the density of the system further decreases. 
It is important to realize that
after freeze-out, one can (by definition) approximately neglect
collisions between particles.  Collective effects related to forward
scattering amplitudes cannot be neglected.  That is, although the
particles no longer scatter, their dispersion relations will not be
given by those in vacuum until the density is further reduced by
continued expansion.

During the expansion, the in-medium sigma mass rises towards its
vacuum value and eventually exceeds the $\pi\pi$ threshold. As the
$\sigma\pi\pi$ coupling is large, the decay proceeds
rapidly. This yields a population of pions with small transverse
momentum, $p_T < m_\pi$.  Because this process occurs after
freeze-out, the pions generated by it do not get a chance to
thermalize.  Thus, the resulting pion spectrum should have a
non-thermal enhancement at low $p_T$ which is largest for
freeze-out near E where the $\sigma$'s are most numerous.

These pions result from the (formerly) long wavelength modes of the
$\sigma$ field, which (unlike $T$ and $\mu$) are expected to fluctuate
at the critical point.  For freeze-out close enough 
to E that the sigma mass at freeze-out is less than $T$,
the thermal fluctuations of the number,
$N_\sigma$, of $\sigma$ particles are determined by the classical
statistics of the field $\sigma$, rather than by Poisson statistics of
particles.  Therefore, ${\langle N_\sigma^2 \rangle - \langle
N_\sigma\rangle^2 \sim \langle N_\sigma\rangle^2}$, rather than
$\langle N_\sigma\rangle$.  Thus, we expect large event-by-event
fluctuations in the multiplicity and distributions of the soft pions:
$N_\pi\approx2N_\sigma$.  Due to critical slowing down,
non-equilibrium effects may further enhance these fluctuations. Thus,
these pions could be detected either directly as an excess in the
$p_T$-spectra at low $p_T$, or via increased event-by-event
fluctuations at low $p_T$, or by an increase in HBT correlations
due to the larger number of pions per phase space cell at low~$p_T$.


To conclude, we propose that by varying control parameters such as
the collision energy and centrality, one may find a window of
parameters for which the $T \mu$ trajectories pass close to the
critical point E. Enhanced critical fluctuations of the $\sigma$ field
and the associated thermodynamic singularities 
lead directly 
to the signatures we propose.  When the freeze-out occurs near
the point E, we predict large non-thermal multiplicity and enhanced
event-by-event fluctuations of the soft pions.  In contrast, the
event-by-event fluctuations in both $T$ and $\mu$, as determined
using pions with $p_T \gtrsim m_\pi$, will be anomalously suppressed. 
Both effects should disappear if
the atomic weight $A$ is very large.  No one of these signatures is
distinctive in isolation and without varying control parameters.
Several of them seen together and seen to turn on and then turn off
again as a control parameter is varied monotonically would constitute a
decisive detection of the critical point.

What would we learn about QCD if such a point is found?  First, we would
learn that there is a genuine critical point in the $T\mu$ plane
in nature. Second, we would learn that $m_s>m_{s3}$ in nature,
and the $\mu=0$ thermal transition is a crossover for physical
quark masses, rather than a first-order phase transition.
Third, the 
experimental 
discovery of the critical end-point E would mean that if the light quark
masses were set to zero, there would be a tricritical point P in
the phase diagram of QCD.

\section{Beyond Monotony}

Most of the signatures which have been proposed as means
of searching for new phases of QCD are expected to 
change monotonically as the collision energy or centrality
is increased.  Once J/$\Psi$'s or $\rho$'s are gone, for example, they
are not expected to come back. 
This makes these signatures susceptible
to mimicry.  The suppression of J/$\Psi$ production or the broadening
of the $\rho$ peak 
can occur at least to some extent in a medium
which is hot and/or dense, but is still on the hadronic side
of the transition region.  The effects of any new phase then
simply augment a previously established monotonic trend, and
may be hard to discern.  This makes non-monotonic signatures
valuable.  One previous example 
arises from the softening of the equation
of state near the phase transition,
whether first or second order or a crossover,
which can yield non-monotonic behavior of observables
related to the lifetime and collective flow of the
plasma.\cite{HS_prl,Rischke,ST}
Here, we have shown how physics unique to the 
region of the critical point in the phase diagram
can yield
non-monotonic behavior of 
the event-by-event fluctuations of suitable observables.
Successful detection 
would demonstrate directly that
freeze-out occurred near the critical point,
and therefore indirectly that
the initial conditions were
well above the transition region.  
Having seen non-monotonic behavior characterizing the critical point,
monotonic observables which can probe the initial conditions directly 
could then be interpreted with greater confidence.

\section*{Acknowledgments}
I am grateful to M. Alford, J. Berges, E. Shuryak, M. Stephanov
and F. Wilczek 
for fruitful collaboration. 
The Minnesota workshop on Continuous Advances in QCD 
was productive and well-named: this contribution is more advanced
than my talk, and some of 
the discussions leading to these advances occurred at the
workshop.
I have had helpful conversations with D. Kim and A. Vainshtein and
with many at the Aspen Center for Physics, where this
contribution was completed, including 
T. Appelquist, R. Birgeneau, N. Evans,
P. Lee, A. Millis,
C. Nayak and M. Schwetz. 
Research supported in part by the A. P. Sloan
Foundation and by the DOE 
through an OJI award and agreement DE-FC02-94ER40818.

\end{document}